# Optical pulling using topologically protected one-way-transport surface-arc waves

Neng Wang[1,2,*,+], Ruo-Yang Zhang[2,*,+], Qinghua Guo[2], Shubo Wang[3], and C. T. Chan[2]

[1]*Institute of Microscale Optoelectronics, Shenzhen University, Shenzhen 518060, China*

[2]*Department of Physics, The Hong Kong University of Science and Technology, Clear Water Bay, Hong Kong, China*

[3]*Department of Physics, City University of Hong Kong, Hong Kong, China*

*Corresponding authors: nwang17@szu.edu.cn, ruoyangzhang@ust.hk

[+] These authors contributed equally

**Abstract:** This paper proposes a new method to achieve robust optical pulling of particles by using an air waveguide sandwiched between two chiral hyperbolic metamaterials. The pulling force is induced by mode conversion between a pair of one-way-transport surface-arc waves supported on the two metamaterial surfaces of the waveguide. The surface arcs bridge the momentum gaps between isolated bulk equifrequency surfaces (EFSs) and are topologically protected by the nontrivial Chern numbers of the EFSs. When an incident surface-arc wave with a relatively small wavenumber $k_{x1}$ is scattered by the particle, a part of its energy is transferred to the other surface-arc mode with $k_{x2}(>k_{x1})$. Because the electromagnetic wave acquires an additional forward momentum from the particle proportional to $(k_{x2}-k_{x1})$ during this process, the particle will always be subjected to an optical pulling force irrespective of its material, shape and size. Since the chiral surface-arc waves are immune to backscattering from local disorders and the metamaterials are isotropic in the *xy* plane, robust optical pulling can be achieved in a curved air waveguide and can go beyond standard optical pulling mechanisms which are limited to pull in a straight-line.

Optical pulling [1-10], which is the counter-intuitive idea of using light to pull a particle towards the source, has attracted great attention recently. Optical pulling forces arise when the incident light transfers a backward linear momentum to the particle via light-matter interaction, which requires the forward scattering to be enhanced and the backward scattering to be suppressed [2-4]. In free space, optical pulling can be achieved by using structured light beams such as a Bessel beam [3-5] or the interference of multiple beams [7]. The technical difficulty in realizing long-range structured beams limits the application of optical pulling. Furthermore, to achieve enhanced forward scattering, multipoles must be simultaneously excited, which renders the optical pulling material- and size-dependent [3, 6, 10]. These problems can be partially solved by changing the background from free space to a waveguide [11-18] or metamaterials [19], such as using the guided waves supported in a double-mode photonic crystal waveguide [16-18]. However, due to the restriction of reciprocity, the material and the shape of the particle have to be carefully optimized in order to suppress the backward scattering.

Using topologically protected one-way-transport surface waves instead of topologically trivial guided waves can enhance the efficacy of optical pulling, since the backscattering of surface waves is completely suppressed for a particle with an arbitrary shape and size and of arbitrary material [20-33]. D. Wang *et al*. first proposed using the chiral surface wave supported in 2D magneto-optical photonic crystals to realize robust optical pulling [35]. However, the practical operating frequency range (gigahertz) of magneto-optical materials rarely overlaps with that of optical manipulation (greater than terahertz), making the optical force negligibly small compared to other forces. In addition, the spatial inhomogeneity of field intensity due to lattice structures can result in local equilibria which work against sustainable optical pulling.

In 3D topological systems possessing Weyl nodes, the topological charges of Weyl points ensure the existence of chiral surface bands whose equifrequency sections form "Fermi arcs" (alternatively called surface arcs for general kinds of waves) that bridge the disconnected equifrequency surfaces (EFSs) of bulk states [31, 32, 34, 36-54]. Such chiral surface-arcs can exist in inversion-symmetry-broken systems respecting time-reversal symmetry [31, 32]. Thanks to the relatively low technological threshold to break inversion symmetry in photonic systems, the surface-arc states have been observed experimentally in regimes ranging from microwave [32, 49, 50] to near-infrared [52], where novel functional transport effects have been realized, such as surface negative refraction [53] and surface-arc-assisted resonant transmission [55]. And in addition to photonic crystals [32, 47, 48, 50], Weyl points and chiral surface arcs can emerge in homogenized metamaterials in the long-wavelength limit, e.g., the chiral hyperbolic metamaterials (CHMs) [31].

In this paper, we propose a method to realize robust long-range optical pulling by using the surface-arc waves. Unlike in 2D magneto-optical photonic crystals, the use of surface-arc waves supported by well-designed CHMs respecting time reversal symmetry can operate at high frequencies to induce a considerable optical force. The linear momentum of the surface wave for a homogeneous medium can be directly determined by the wave vector, unlike the ambiguous relation between the linear momentum of a Bloch state and its Bloch wave vector in periodic systems [35, 55], and hence the mechanism of momentum transfer is much clearer in homogenized metamaterials than in photonic crystals. Moreover, the uniform field distribution of the surface-arc wave along the propagating direction avoids the local equilibria induced by the intensity gradient force, thereby enabling consistent pulling of the particle towards the source.

A generic class of homogenized CHMs can be described by a Drude-Lorentz model with chiral magnetoelectric coupling [32,50,46,48]. Here, we assume that the metamaterials have a Drude-like dispersion in the *z* direction, and isotropic Lorentz resonances in the *xy* subspace. Additionally, the aligned electric and magnetic dipoles are chirally coupled in the *xy* plane. From the microscopic model (ignoring damping for simplicity) [58], we arrive at a macroscopic effective medium depicted by the constitutive relation $\mathbf{D} = \varepsilon_0 \vec{\varepsilon}(\omega)\mathbf{E} + \frac{i}{c}\vec{\kappa}(\omega)\mathbf{H}$, $\mathbf{B} = \mu_0 \vec{\mu}(\omega)\mathbf{H} - \frac{i}{c}\vec{\kappa}(\omega)\mathbf{E}$, and the relative constitutive tensors

$$\vec{\varepsilon}(\omega) = \left(\varepsilon_\infty + \frac{\alpha\omega_0^2}{\omega_0^2 - \omega^2}\right)\vec{I}_{xy} + \left(1 - \frac{\omega_p^2}{\omega^2}\right)\hat{z}\hat{z},$$

$$\vec{\mu}(\omega) = \left(1 + \frac{\beta\omega^2}{\omega_0^2 - \omega^2}\right)\vec{I}_{xy} + \hat{z}\hat{z}, \quad (1)$$

$$\vec{\kappa}(\omega) = \kappa(\omega)\vec{I}_{xy} = \pm\frac{\sqrt{\alpha\beta}\,\omega_0\omega}{\omega_0^2 - \omega^2}\vec{I}_{xy},$$

where $\omega_0, \omega_p$ are the Lorentz resonance and plasma frequencies, respectively, $\varepsilon_\infty$ denotes the asymptotic relative permittivity in *xy*-plane at high frequencies, and $\alpha, \beta$ are two positive parameters characterizing the strengths of Lorentz oscillation and chiral coupling, respectively. Since Eq. (1) is deduced from the microscopic Hermitian dynamics [58], the eigen-frequency $\omega(\mathbf{k})$ is ensured to be real, where $\mathbf{k}$ is the plane wave vector of the eigen-mode. The band structure is plotted in Fig. 1(a). As the chiral coupling breaks spatial inversion symmetry while preserves time-reversal symmetry, two pairs of Weyl points emerge symmetrically on the line of $k_x = k_y = 0$ at the plasma frequency $\omega_p$. If the frequency is in between $\max\{\omega_0\sqrt{1+\alpha/\varepsilon_\infty}, \omega_0/\sqrt{1-\beta}\}$ and $\omega_p$, Eq. (1) gives a Type-I hyperbolic medium ($\varepsilon_{xx} = \varepsilon_{yy} > 0, \varepsilon_{zz} < 0, \mu_{xx} = \mu_{yy} > 0$) [57] with chiral coupling $\vec{\kappa}$. Its bulk EFSs consist of a closed spheroid and a hyperboloid of two sheets, separated by two complete $k_z$-momentum gaps. The spheroidal EFS and each hyperbolic sheet carry topological charges (i.e. Chern numbers) $\mathrm{Ch} = -2\,\mathrm{sgn}(\kappa)$ and $\mathrm{Ch} = \mathrm{sgn}(\kappa)$, respectively. We note that such CHMs can be realized using metallic helical structures, see details in the Supplemental Materials (SM) [58].

When the metamaterial is in contact with air at a flat surface parallel to the *z* axis, the nontrivial Chern numbers of the EFSs indicates the existence of surface states confined on that surface which disperse as open arcs bridging the projected EFSs [31], according to the correspondence between bulk EFSs and surface arcs [41,42]. As shown in Fig. 1(b), since the spheroidal EFS has $\mathrm{Ch} = -2\,\mathrm{sgn}(\kappa)$, two surface arcs tangentially attaching to the its projection stretch across the $k_z$ gaps and connect with the upper and lower charge-1 hyperbolic sheets, respectively. The gapless surface arcs at different frequencies sweep out two surface bands (colored in yellow) filling the bulk bandgaps, as shown in Fig. 1(a). Consider a virtual equifrequency loop enclosing the projected spheroidal EFS and inevitably intersecting with the two surface arcs, where the orientation of the loop is counterclockwise viewed from the top of the truncated surface. At an intersection point $p$, we define the chirality of the surface arc as $\chi_p = \text{sgn}\left[d\mathbf{k}_{loop} \cdot \mathbf{v}_g\right]$ [42], where $d\mathbf{k}_{loop}$ is the positive tangential vector of the loop, and $\mathbf{v}_g$, normal to the surface arc, denotes the group velocity of the surface-arc state at $p$. The summation of the chirality of all intersections equals the Chern number of the EFS enclosed by the loop, namely $\sum_p \chi_p = \text{Ch} = -2\,\text{sgn}(\kappa)$. For a rectangular loop in Fig. 1(b), the two horizontal paths along $k_z = \pm k_{z0}$ intersect with the two surface arcs at $p_+, p_-$, respectively. Due to time-reversal symmetry, the two intersections must have the same chirality, i.e. $\chi_{p_+} = \chi_{p_-} = -\text{sgn}(\kappa)$. Therefore, the longitudinal (along the $x$ direction) propagating directions of the surface-arc states are determined by the topological charge of the EFS. In what follows, we illustrate how to realize robust optical pulling using such chiral surface-arc states.

The configuration of the optical pulling system is schematically shown in Fig. 2 (a), where an air gap is sandwiched between two lossless CHMs with constitutive parameters. The particle can move in the channel along the $x$ direction, as indicated in Fig. 2(a). For the two metamaterial surfaces facing the air gap, each one supports a topologically protected chiral surface arc in the upper $k_z$ gap between the bulk EFSs, and the corresponding surface-arc states propagate unidirectionally. In Fig. 2(e), the red solid (blue dashed) line depicts the surface arc on the interface between the upper (lower) CHM and air, which we refer to arc 1 (arc 2), and we call the surface states on the two arcs state 1 and state 2, respectively. A line source (red star in Fig. 2(a)) is located near the left surface of the upper lossless CHM to excite state 1. To prevent the surface waves from returning to the starting point after circling the CHMs, we attach an absorbing layer to each lossless CHM, as shown in Fig. 2(a). For the chosen $k_z = 1.3k_0$, where $k_0$ is the vacuum wavenumber, the longitudinal wavenumber $k_{x2}$ of state 2 is less negative than $k_{x1}$ of state 1, which can be seen from Fig. 2(e). The group velocities of the surface-arc states are normal to the corresponding surface arcs as marked by the arrows in Fig. 2(e). At $k_z = 1.3k_0$, both the two surface-arc states are propagating from left to right inside the air gap. Note that both $k_{x1}$ and $k_{x2}$ are negative, therefore, the $x$ components of the phase and group velocities of each state are in the opposite directions.

Consider a round particle located in the air gap. Since state 2 cannot be exited due to the mismatch of $k_x$, the incident surface-arc wave will be confined on the surface of the upper CHM before scattered by the particle, as shown by the simulation result in Fig. 2(b) obtained using the commercial package COMSOL [59]. When the surface-arc wave is scattered by the particle, plane wave components of various $k_x$ are generated around the particle (acting as a secondary line source) and state 2 is excited by the component with $k_x = k_{x2}$. Since only the two modes with $k_x = k_{x1}$ and $k_x = k_{x2}$ can propagate, all the optical energy will be distributed to states 1 and 2. Due to the absence of backward channel, the incident wave cannot be reflected by the particle, independent of its material content, size and shape. Therefore, even if we replace the round particle by a triangular one, the backward scattering is still totally suppressed, as shown in Fig. 2(c).

The linear momentum of a surface mode supported by a homogeneous medium is proportional to its propagating wave vector [60]. Interestingly, states 1 and 2 have backward momentum while they are propagating forward along the $x$ direction. As long as a part of the incident surface-arc wave is transferred into the other one with a less negative $k_x$ incurred by the scattering, the total forward momentum of the light is increased. Consequently, the particle will gain a backward momentum from the light by momentum conservation and will be pulled backward. We emphasize that, no matter how much energy of the incident surface-arc wave is transferred into the other mode due to the particle scattering, the total forward momentum of light will always increase. The optical pulling is hence robust irrespective of the particle's material, size and shape. The system parameters only affect the conversion efficiency from state 1 to state 2 and hence the magnitude of the optical pulling force. In Figs. 3(a) and (b), we show the longitudinal optical forces $f_x$ of (a) a round and (b) a triangular dielectric particle located in the air gap as functions of their relative permittivities and dimensionless size parameters. It is seen that the longitudinal optical force is always negative. The strong optical pulling force denoted by the blue regions is attributed to the multipolar resonances of the particles.

According to momentum conservation, the pulling force is proportional to the wavenumber difference between the two surface-arc states $\Delta k_x = k_{x2} - k_{x1}$. This gives us a way to adjust the strength of the pulling force. By changing the relative permittivity tensor of the lower CHM to $\vec{\varepsilon}_2'$, the value of $\Delta k_x$ is reduced (the momentum gap between the red solid and blue dashed lines in Fig. 2(f) is smaller than that in Fig. 2(e)). If we further change the sign of the chiral coupling to be $\kappa_2' = -\kappa_2$, the chirality of arc 2 (green dashed line) will also be changed, and the longitudinal wavenumber $k_{x2}$ will become positive, leading to larger $\Delta k_x$ (gap between the red solid and green dashed lines in Fig. 2(f)) than the case of $k_{x2} < 0$. The curves in Fig. 3(c) illustrates the longitudinal optical force $f_x$ at $k_z = 1.3 k_0$ as a function of the relative permittivity $\varepsilon_r$ for three different $\Delta k_x$, showing that the smaller $\Delta k_x$ is, the weaker the optical pulling force will be. Interestingly, when state 2 lies on the green dashed line in Fig. 2(f), the incident light will be reflected by the particle, as shown by the full-wave simulation in Fig. 2(d). Counterintuitively, the optical force is still pulling since $\Delta k_x > 0$.

The optical pulling force is proportional to the total energy flux transferred from state 1 to state 2 during the scattering, which can be quantitatively verified by studying the field intensity of state 1 before and after it is scattered. Here we consider two points labeled as "A" and "B" in Fig. 2(b), which are slightly below the upper surface of the air gap with the same distance. Since the points are far away from the lower surface of the air gap, the electric fields $\mathbf{E}_A$ and $\mathbf{E}_B$ at the two points are almost totally contributed by state 1. Note that the shape of field intensity distribution of state 1 is independent of the $x$ coordinate, and the total energy fluxes of state 1 before and after the scattering are proportional to the field intensities $|\mathbf{E}_A|^2$ and $|\mathbf{E}_B|^2$, respectively. In Fig. 3(c), we show the longitudinal optical force $f_x$ and the normalized scattered energy

flux, $I_x = (|\mathbf{E}_B|^2 - |\mathbf{E}_A|^2)/|\mathbf{E}_A|^2$, as functions of the permittivity by the black solid line and black circles, respectively. The circles fall right on the line, indicating that the optical pulling force is proportional to $I_x$. Also, we numerically computed $f_x$ normalized by $I_x$ for different examples obtained by adjusting the constitutive parameters of the lower CHM. The result in Fig. 3(d) shows that the normalized force is linearly dependent on $\Delta k_x$.

Because the CHM is isotropic in the *xy* plane, the chiral surface arcs on any truncated surface parallel to *z*-axis have the same dispersion, and the surface-arc waves can travel around corners without any reflection. Thus, a particle can be pulled continuously in a curved air gap. As a typical example, we consider that the air waveguide has a 90º bend, as shown in Fig. 4. The particle is first subjected to an upward optical force in the vertical region until reaching the corner region, then move into the horizontal region directly or after elastic collisions with the modified corner boundaries, see details in the SM [58]. After that, it will be pulled towards the left end. Although only straight and 90º bending waveguides are shown here, we emphasize that the particle can be pulled along a complex trajectory defined by a meandering waveguide [58].

In summary, we proposed a scheme to realize robust and long-range optical pulling using two topologically protected chiral surface arcs on two facing CHMs. The optical force on the particle originates from the momentum transfer between the two surface-arc modes during the scattering process. We can realize optical pulling or switch it to optical pushing by simply selecting the launched state, irrespective of the particle's material, size and shape. The advantage of employing a metamaterial is its homogeneity in the long wave limit. Benefiting from the isotropy of the metamaterial in the *xy* plane, we can pull the particle along an arbitrary 2D trajectory by designing the shape of the surface-wave waveguide. With the advance in nano-engineering technology for realizing hyperbolic and chiral metamaterials at the working wavelengths of optical forces [57,61], our robust optical pulling scheme can provide an additional tool for the optical micromanipulations of matter.

**Acknowledgements:** This work is supported by National Natural Science Foundation of China (NSFC) through No. 11904237 and the Research Grants Council of the Hong Kong Special Administrative Region through grant Nos. AoE/P-02/12, 16303119 and CityU 11306019.

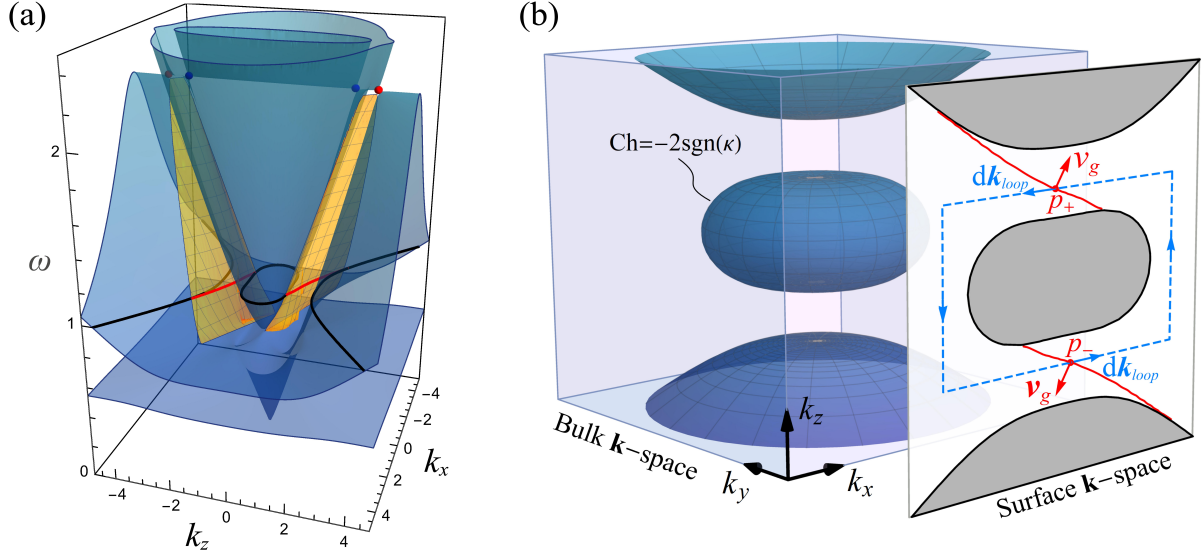

**Fig**. 1. (a) Band structure of the CHM given by Eq. (1) with the parameters $\omega_0 = 0.5, \omega_p = \sqrt{5}, \alpha = 1.2, \beta = 0.375$, where four Weyl points with charges $\text{sgn}(\kappa)$ (red nodes) and $-\text{sgn}(\kappa)$ (blue nodes) are located at the plasma frequency $\omega_p$. The yellow sheets denote the bands of chiral surface states on the interface between the metamaterial and air with the surface normal $\mathbf{n} = -\hat{y}$. (b) Bulk EFS at $\omega = 1$ (corresponding to the black section line in (a)), comprised of a spheroid with $\text{Ch} = -2\text{sgn}(\kappa)$ and two hyperbolic sheets. Two surface-arcs (red lines) bridge disjoint projected bulk EFSs, intersecting a rectangular loop (blue dashed line) at $p_+$, $p_-$. Counting the chirality $\chi = \text{sgn}(\mathbf{v}_g \cdot d\mathbf{k}_{loop})$ at each intersection, the summation equals the Chern number of the EFS enclosed by the loop.

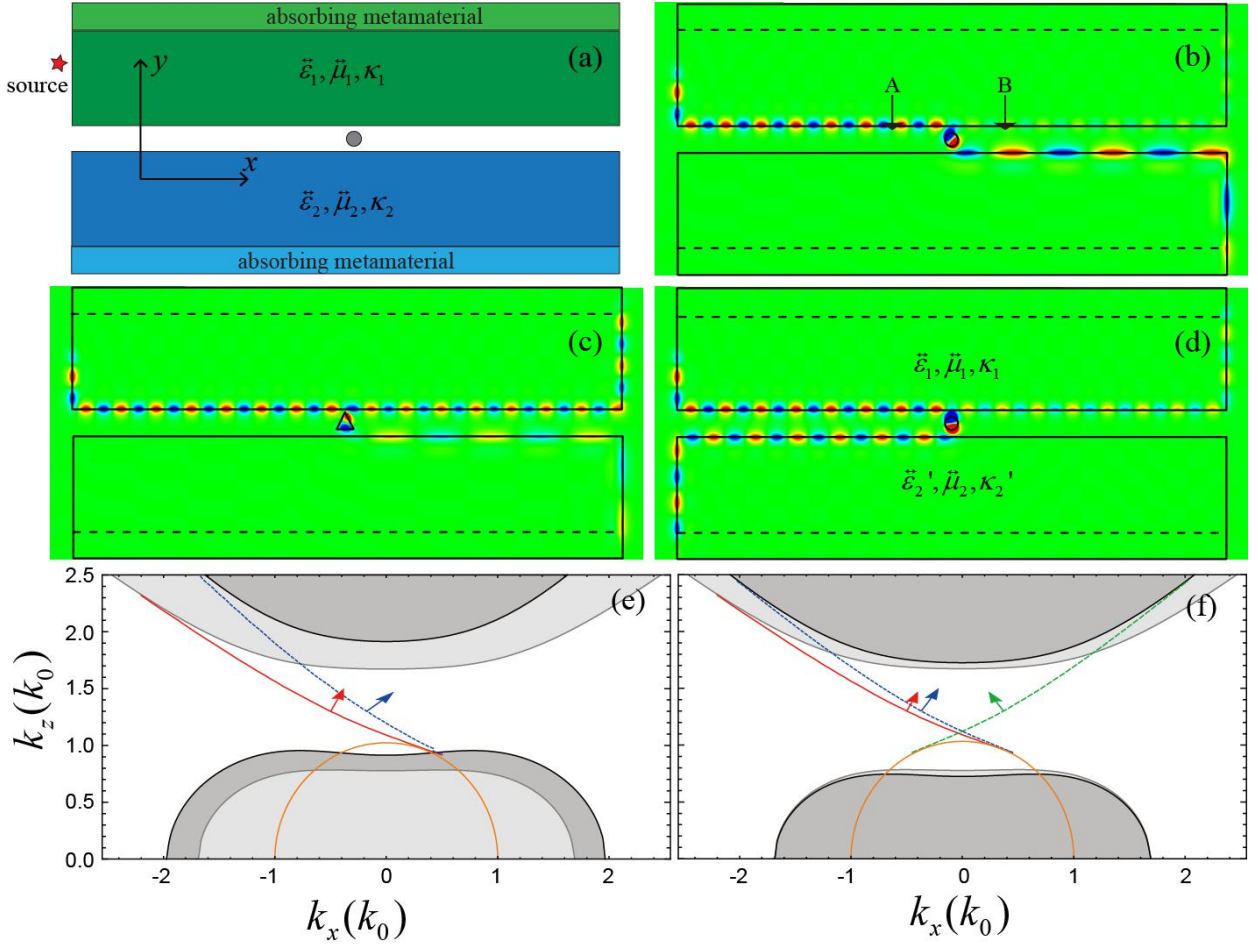

**Fig 2.**(a) Configuration of the optical pulling system. The constitutive parameters of the upper CHM are $\vec{\varepsilon}_1 = \text{diag}(3,3,-4)$, $\vec{\mu}_1 = \text{diag}(0.5,0.5,1)$, $\kappa_1 = 0.45$ and the relative permittivity of the dielectric particle in the air gap is $\varepsilon_r = 3$. (b)-(c) The constitutive parameters of the lower CHM are $\vec{\varepsilon}_2 = \text{diag}(4,4,-3)$, $\vec{\mu}_2 = \text{diag}(0.5,0.5,1)$, $\kappa_2 = -0.5$. The points "A" and "B" marked by the black arrows in (b) are selected to measure the fields before and after the scattering. (d) The constitutive parameters of the lower CHM are $\vec{\varepsilon}'_2 = \text{diag}(3,3,-3)$, $\vec{\mu}_1$, $\kappa'_2 = 0.5$. (e) and (f) Dispersion of state 1 (red solid lines) and state 2 (blue dashed and green dashed lines). The arrows mark the directions of group velocities. The shaded regions correspond to the bulk states of CHM and the orange solid lines denote the EFS of air. (e) The upper and lower CHMs are the same as those used in (a)-(c). (f) The upper CHM takes the same parameters as in (a)-(c), for the blue dashed line the lower CHM has the relative permittivity tensor $\vec{\varepsilon}'_2$ and chirality $\kappa_2$, while for the green dashed line the lower CHM has the relative permittivity tensor $\ddot{\varepsilon}'_2$ and chirality $\kappa_2'$.

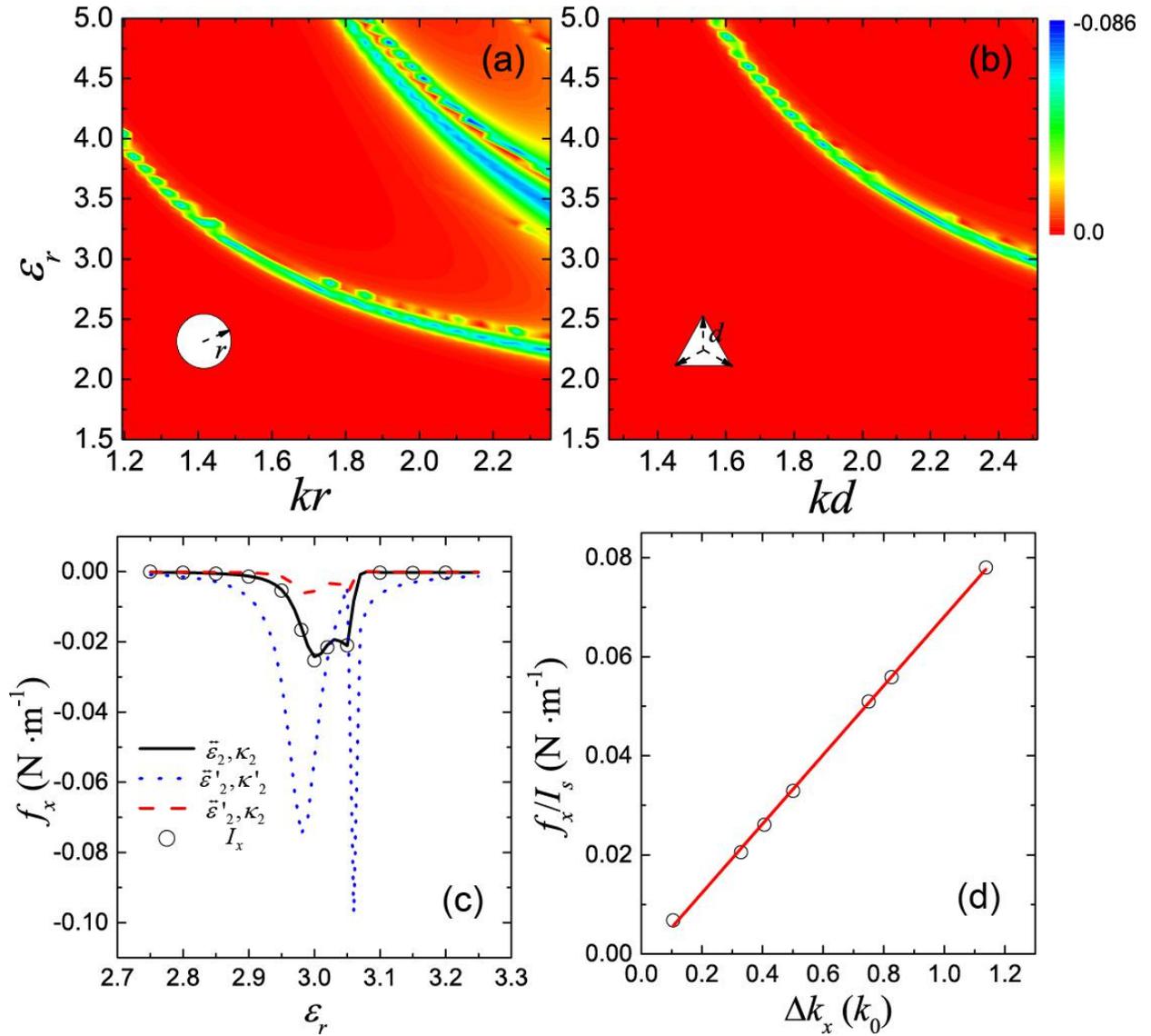

**Fig. 3**. The optical force along the $x$ direction $f_x$ as a function of the dimensionless size parameter (a) $k_0 r$ and (b) $k_0 d$ and the relative permittivity $\varepsilon_r$ of the (a) round and (b) triangular

particle. The unit of the force is $N \cdot m^{-1}$ and the amplitude of the source current is 1A. The out-of-plane vacuum wavenumber is $k_z = 1.3k_0$. (c) $f_x$ acting on the round particle ($k_0 r = 0.5\pi$) as a function of $\varepsilon_r$ for three different lower CHMs with relative permittivity tensors and chirality ($\vec{\varepsilon}_2, \kappa_2$) (black solid), ($\vec{\varepsilon}_2', \kappa_2'$) (blue dots) and ($\vec{\varepsilon}_2', \kappa_2$) (green dashed). The circles display the normalized scattered energy flux $I_x$. (b) Linear fitting of $f_x/I_x$ (red line) as a function of the wavenumber difference $\Delta k_x = k_{x2} - k_{x1}$.

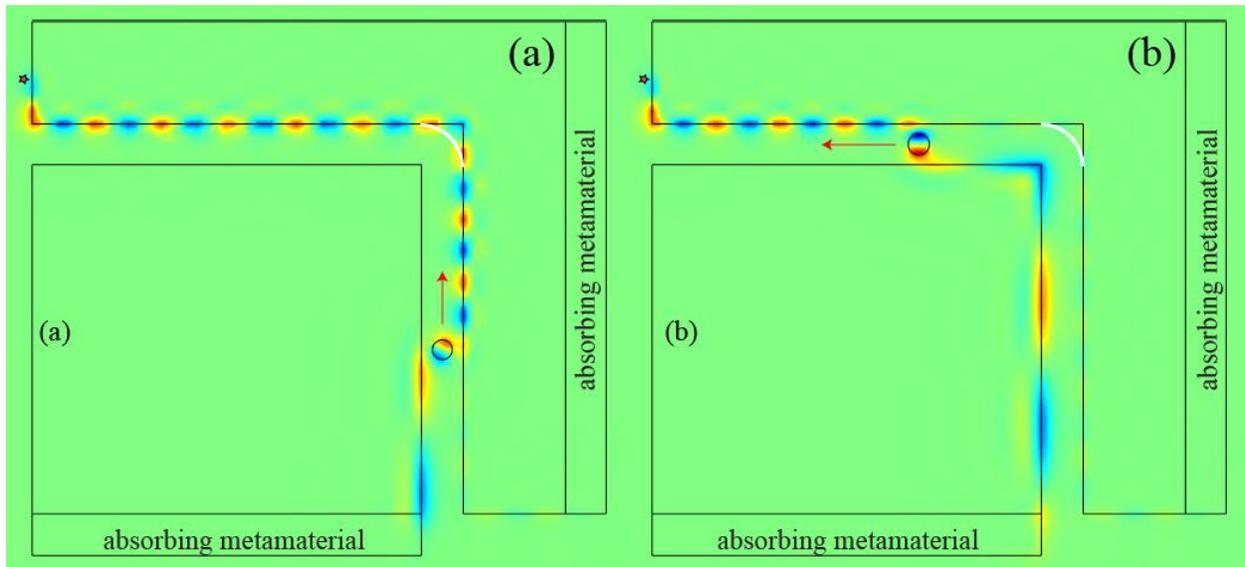

**Fig. 4.** The profile of surface-arc waves inside the air waveguide with a 90º bend. The white arc denotes a thin and transparent barrier. The arrows show the optical force directions.